\documentclass[12pt,A4]{article}
\usepackage{graphics}
\usepackage{color}
\begin{document}
\begin{center}

{\large\bf $\pi^+\pi^+$ and $\pi^+\pi^-$ colliding in noncommutative
space }

\vskip 1cm
 Wang Jianhua$^{1}$~~Li Kang$^{2}$~~Sayipjamal Dulat$^{3}$~~Yuan Yi$^{1}$~~Ma Kai$^{3}$
\\\vskip 1cm

{\it\small 1.~~ Shaanxi University  of Technology, Hanzhong, 723001,
Shaanxi, China

2.~~ Hangzhou Normal University, Hangzhou, 310036, Zhejiang, China

3.~~ Xinjiang University, Urumqi, 830046, Xinjiang, China
\\}
\end{center}

\begin{abstract}
\noindent  By studying the scattering process of scalar particle
pion on the noncommutative scalar quantum electrodynamics, the
non-commutative amendment of differential scattering cross-section
is found, which is dependent of polar-angle and the results are significantly
different from that in the commutative scalar quantum
electrodynamics, particularly when $\cos\theta\sim \pm 1$.
The non-commutativity of space is expected to be explored at around
$\Lambda_{NC}\sim$TeV.

\noindent PACS number(s): 11.10.Nx, 25.80.Dj, 13.85.-t.

\end{abstract}

\section{Introduction}
The emergence of the noncommutative geometry with a neutral way in
string theory/(M-Theory) in a definite limit not only makes us
effectively analyze the duality, BPS state and D-brane dynamics, but
also causes a revolution in the whole physical theory because a
fresh idea -- noncommutative spacetime was put forward$^{[1-10]}$.
This means that in the Planck-scale the space-time coordinates are
no more commutative and they will satisfy an uncertain-relation ,
the space-time point loses its original sense
and the geometry which describes the original physical phenomenons
is not consistent with the new physics in this space-time area.
Therefore, it is necessary to have a new space-time geometry
--noncommutative space-time geometry to describe the
gravitation$^{[11-23]}$. Thus, there have been a lot of delightful
achievements in the noncommutative field theory, the topological
phase and the correction of noncommutative energy levels in recent
years$^{[24-28]}$. In addition, Refs.[29,30] have studied the Wigner
function in noncommutative space.

In this paper we focus on the scattering process of scalar particle
pion on the noncommutative scalar quantum electrodynamics(NCSQED). In
section 2, we mainly review the Weyl-Moyal product in noncommutative
space. In section 3 we give the Feynman rules for NCSQED and show
that there are some differences between the quadric kinematic terms
in NCSQED and in noncommutative electrodynamics(NCED). In section 4 and 5 we discuss the elastic
scattering processes of $\pi^+\pi^+$ and $\pi^+\pi^-$ respectively.
The annihilation of the scalar particles is studied in section 6,
and there are discussions on three and four-point photon functions
($\theta^{0i}\neq0$). Conclusion and further discussion are given in
the last section.

\section{The Weyl-Moyal product in noncommutative space}
In this part we mainly review the Weyl-Moyal product in
noncommutative space. As is known that the study of noncommutative
field theory arose from the pioneering work by Snyder$^{[1]}$. In
recent years, with the development of string theory,  it has been paid
more attention and great progresses have been made. For example,
noncommutativity of space is an important characteristic of D-brane
dynamics at low energy limit$^{[2,3]}$. What's more, interests are
increasing in possible physically observable consequences related to
non-commutativity of space, which is expected to justify
noncommutative field theory and noncommutative quantum mechanics
with the references of seminal works$^{[4-10]}$. In general,
noncommutativity relation of space-time can be described as:
\begin{equation}\label{1}
    [\hat{x}_{\mu},\hat{x}_{\mu}]=i\theta_{\mu\nu}=\frac{ic_{\mu\nu}}{\Lambda^2_{\rm{NC}}}
\end{equation}
where $\theta_{\mu\nu} $ is an anti-commutation parameter and of
dimension $[L]^2$, $\Lambda_{\rm{NC}}$ is the scale, in this scale the
space-time is no longer commutative. The action for field theory in
noncommutative space is then obtained by using the Weyl-Moyal
correspondence$^{[11]}$. Thus, in order to find the noncommutative
action the usual product of field should be replaced by the
star-product,
\begin{equation}\label{2}
    (f\ast g)(x)=\exp{\left[\frac{i}{2}\theta_{\mu\nu}\partial_{x_{\mu}}\partial_{x_{\nu}}\right]}
    f(x)g(y)|_{x=y}~,
\end{equation}
in which $f$ and $g$ are two arbitrary infinitely differentiable
functions on $R^{3+1}$. This product has an important character,
\begin{equation}\label{3}
    \int d^4x (f\star g)(x)=\int d^4x (g\star f)(x)=\int d^4x (f\cdot
    g)(x).
\end{equation}
This means that if the free Lagrangian only include quadric field
variable, we can construct the same Fock space as in communicative
space to perform the perturbative evaluation$^{[11]}$. In the case
of $\theta^{0i}=0$ , it has been shown that noncommutative $\psi^4$
theory up to two loops$^{[11,12]}$ and noncommutative QED up to the
one loop$^{[13,14]}$ are renormalizable. While($\theta^{0i}\neq0$),
it has been shown that the Noncommutative field theory is not
unitary. So, the theory is not appealing$^{[15]}$.

Noncommutative action can also be obtained according to
Seiberg-Witten(SW) map$^{[16]}$. Using this SW-map, Calmet and
others first constructed a model with noncommutative gauge
invariance which was close to the usual standard model and is known
as the minimal NCSM(mNCSM). They also listed several Feynman
rules$^{[17]}$. Although many phenomenological studies$^{[18-22]}$
have been made to unravel several interesting features of this
mNCSM, Raimar Wulkenhaar has proved that ``\emph{Noncommutative
scalar QED cannot be renormalized by means of Seiberg-Witten
expansion.}"$^{[23]}$. He has provided some ideas for how the
Seiberg-Witten expansion can be used as a computational technique to
treat one-loop divergences of the $\theta$-unexpanded noncommutative
QED.

\section{Feynman rules for NCSQED}
With the revision of Weyl-Moyal product we provide Feynman rules for
NCSQED in this section. Assuming Raimar Wulkenhaar's word $^{[10]}$
also makes sense for noncommutative scalar quantum electrodynamics,
we study the full electromagnetic scattering of noncommutative
scalar quantum electrodynamics. The full
lagrangian in noncommutative field theory can be obtained  by replacing the usual
product with the star product. So, the lagrangian for the scalar
noncommutative can be written as
\begin{equation}\label{4}
    {\mathcal {L}}=-\frac{1}{4}F_{\mu\nu}\star F^{\mu\nu}
    +(D_{\mu}\phi)^*\star(D_{\mu}\phi),
    -m^2\phi^*\star\phi,
\end{equation}
which is invariant under the following transformations
\begin{equation}\label{5}
    \phi(x,\theta)\rightarrow \phi'(x,\theta)=U\star \phi(x,\theta),
\end{equation}
\begin{equation}\label{6}
    A_{\mu}(x,\theta)\rightarrow A_{\mu}'(x,\theta)=U\star
    A_{\mu}(x,\theta)\star U^{-1}+\frac{i}{e}U\star
    \partial_{\mu}U^{-1}
\end{equation}
where $U=(e^{i\alpha})_{\star}$ and
\begin{equation}\label{7}
    D_{\mu}\phi=\partial_{\mu}\phi-ieA_{\mu}\star\phi ,
\end{equation}
\begin{equation}\label{8}
    F_{\mu\nu}
    =\partial_{\mu}A_{\nu}-\partial_{\nu}A_{\mu}-ie(A_{\mu}\star
    A_{\nu}-A_{\nu}\star A_{\mu}).
\end{equation}
Obviously, besides the photon's 2-point function(which is identical to the
ordinary case because its forms remain unchanged (see Eq.(3))), the
photon's 3- and 4-point functions are now transparent. After transforming
these new interactions into momentum space, the vertices pick up
additional phase factors which are dependent upon the momenta
flowing through the vertices. These kinematic phases play an
important role in the colliding process of the noncommutative field
theory. Now that All Feynman rules are in Ref.(5), here we only
mention Feynman rules which will be used in this paper. The photon's
vertices are same as of the NCQED, and Feynman rules for the
photon's 3-point vertex are shown explicitly in Fig.1, where
$p\wedge k=p\cdot\theta\cdot k$ is used as notation.
\begin{figure}[h]
\centering
\includegraphics{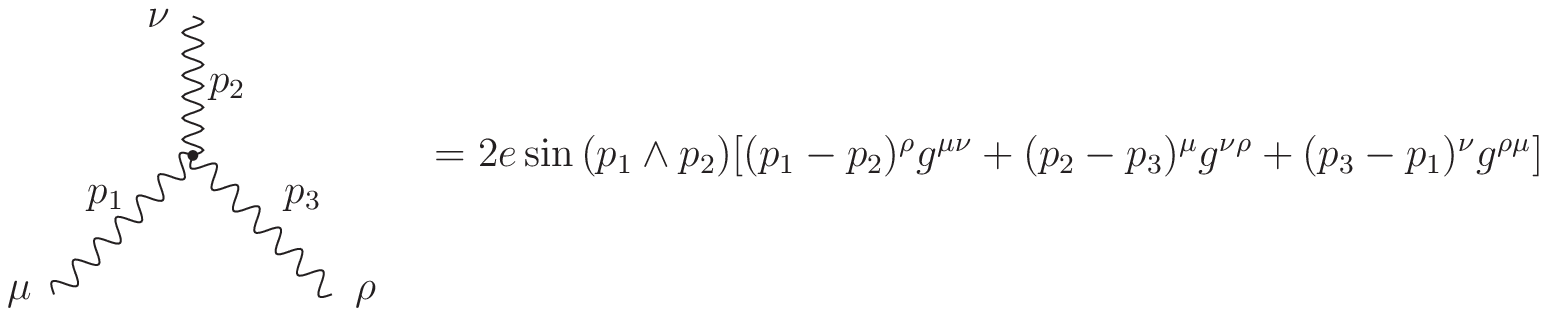}
\caption{The Feynman rules for 3-point vertex} \label{1}
\end{figure}

The interacting Lagrangian of scalar particles and photons can be
written as
\begin{equation}\label{9}
    {\mathcal {L}}_{S-B}=ieA^{\mu}(\phi^*\star\partial_{\mu}\phi
    -\phi\star\partial_{\mu}\phi^*)-e^2 A^{\nu}\star A_{\nu}
    \star\phi^*\star\phi.
\end{equation}
Obviously, different from the 3- and 4-point functions of photons,
the very modification of noncommutativity of space-time is the
additional kinematic phase factor. Thus, according to Eq.(9) we can
easily obtain Feynman rules for these interactions in Fig.2. (It
should be noted that there are two ways to calculate Feynman rules
for the second term of Eq.(8), so an average should be made.)
\begin{figure}[h]
\centering
\includegraphics{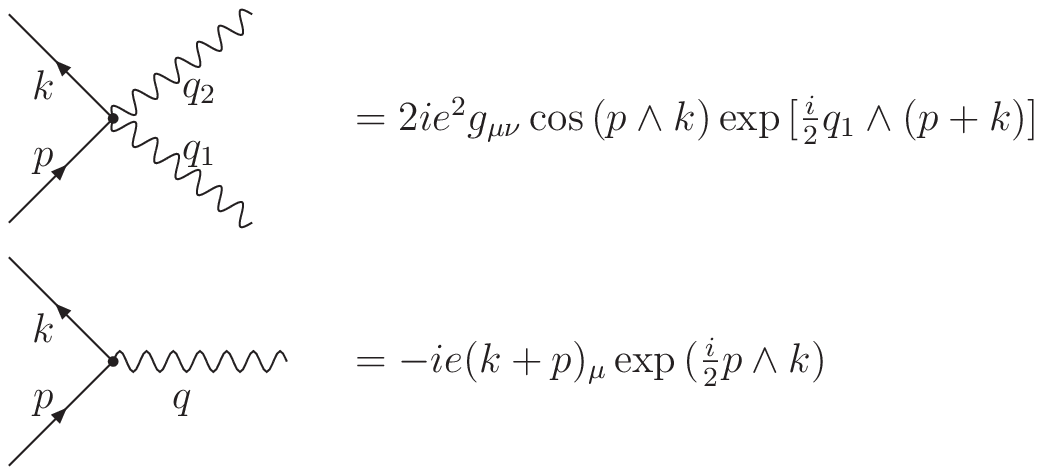}
\caption{The Feynman rules for interactions of scalar particles to
photon} \label{2}
\end{figure}

\section{$\pi^+\pi^+$ elastic scattering in NCSQED}
As the M\"{o}ller and Bhabha scattering in the non-commutative
electrodynamics $^{[6]}$, the reaction $\pi^+\pi^+ \rightarrow
\pi^+\pi^+$ and $\pi^+\pi^- \rightarrow \pi^+\pi^-$ is a simple
model which can be used to test the non-commutativity of space in
the NCSED. However, there are some differences from the NCQED due to
the quadric kinematic terms in the Lagrangian. This can be realized
explicitly form the interaction Lagrangian (8), as well as from the
Feynman rules in Fig.2. These differences are expected to guide a
new approach to test the noncommutativity of space during the
colliding process of high-energy particles.

Let us first consider the elastic $\pi^+\pi^+$ colliding, whose
Feynman diagrams are displayed in Fig.3.
\begin{figure}[h]
\centering
\includegraphics{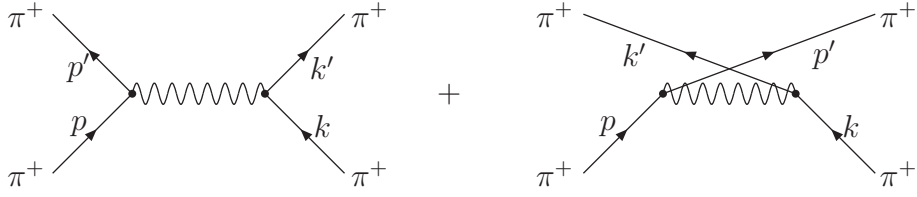}
\caption{The scattering of $\pi^+ \pi^+$} \label{3}
\end{figure}
In this case, kinematic phase corresponding to NC modifications
appears in each vertex. Following Feynman rules in Fig.2 and
momentum labels in Fig.3, we can get the invariant matrix element of
the scattering process
$$
    i{\mathcal {M}}_{++}=-ie^2 \exp{\left[{i\over 2}(p\wedge p'+k\wedge k')\right]}
    (p+p')^{\alpha}(k+k')^{\beta}\frac{g_{\alpha\beta}}{q^2}
    ~~~~~~~~~
$$
\begin{equation}\label{10}
    -ie^2 \exp{\left[{i\over 2}(p\wedge k'+k\wedge p')\right]}
    (p+k')^{\alpha}(k+p')^{\beta}\frac{g_{\alpha\beta}}{q^2}.
\end{equation}
Obviously, $t$- and $u$-channel interference terms hold a factor
$\cos{\delta}$ where $\delta$ is the phase difference of $t$ and
$u$-channel,
\begin{equation}\label{11}
    \delta_{++}={1\over 2}(p\wedge p'+k\wedge k')-{1\over 2}(p\wedge k'+k\wedge
    p').
\end{equation}
It should be noted that all terms involving time-space
noncommutativity have been dropped out of the expression of
$\delta_{++}$ because this theory is not unitary and hence, as a
field theory, it is not appealing$^{[10]}$. All other effecting
quantities are shown in the following formulas (in which
center-of-mass frame and Mandelstam variables are used.),
\begin{eqnarray}
  p\wedge p' &=& \frac{s}{4}(\theta_{12}\sin{\theta}\cos{\phi}
  -\theta_{31}\sin{\theta}\sin{\phi}), \\
  p\wedge k' &=& \frac{s}{4}(-\theta_{12}\sin{\theta}\cos{\phi}
  +\theta_{31}\sin{\theta}\sin{\phi}), \\
  k\wedge p' &=& \frac{s}{4}(-\theta_{12}\sin{\theta}\cos{\phi}
  +\theta_{31}\sin{\theta}\sin{\phi}), \\
  k\wedge k' &=& \frac{s}{4}(\theta_{12}\sin{\theta}\cos{\phi}
  -\theta_{31}\sin{\theta}\sin{\phi}).
\end{eqnarray}
Using these formulas we can simplify the expression of $\delta$
\begin{equation}\label{16}
    \delta_{++}=-\sqrt{ut}(\theta_{12}\cos{\phi}-\theta_{31}\sin{\phi}).
\end{equation}
In addition, as we take the limit $\Lambda_{NC}\rightarrow \infty$,
$\cos{\delta}=1$ the ordinary result is recovered. Using the
Mandelstam variables we can obtain the squared invariant matrix
element after a simple calculation
\begin{equation}\label{17}
    |{\mathcal {M}}_{++}|^2=\bigg(\frac{4\pi\alpha}{t}\bigg)^2\bigg[(s-u)^2
    +(s-t)^2+2\cos{\delta}(s-u)(s-t)\bigg].
\end{equation}
 While, in the high-energy limit ($m\rightarrow 0$) the Mandelstam variables have the
following simple forms
\begin{eqnarray}
    s &=& (2E)^2=E^2_{\rm{CM}} \\
    u &\approx & -2E^2(1+\cos{\theta}) \\
    t &\approx& -2E^2(1-\cos{\theta}).
\end{eqnarray}
Here the note $E$ is the energy of the incoming particle. What's
more, the differential cross section can be deduced
\begin{equation}\label{21}
    \bigg(\frac{\rm{d}\sigma_{++}}{\rm{d}\Omega}\bigg)
    _{\rm{CM}}=\frac{|{\mathcal {M}}_{++}|^2}{64\pi^2s^2}.
\end{equation}

In numerical analysis, we set the machine energy for
$\Lambda_{NC}=\sqrt{s}=1.2~\rm{TeV}$. We only consider one
non-vanishing value of $\theta_{ij}$ and set its magnitude for
$1/\Lambda_{NC}^2$. Furhter, for simplicity we just consider the
case $\theta_{31}=\theta_{12}=1/\Lambda^2_{NC}$. In this way, when
$\phi=\pi/4$, the results will be obviously same as in the
commutative case. In Fig.(4), we plot the angular distribution of
the ratio between the noncommutative modification and the
differential cross section, $(\rm{d}\sigma-\rm{d}\sigma_{NC})/
\rm{d}\Omega$. We find that the ratio vanish at $\theta=0,\pi$,
i.e.there is no noncommutative modification in this case.
\begin{figure}[h]
\centering
\includegraphics{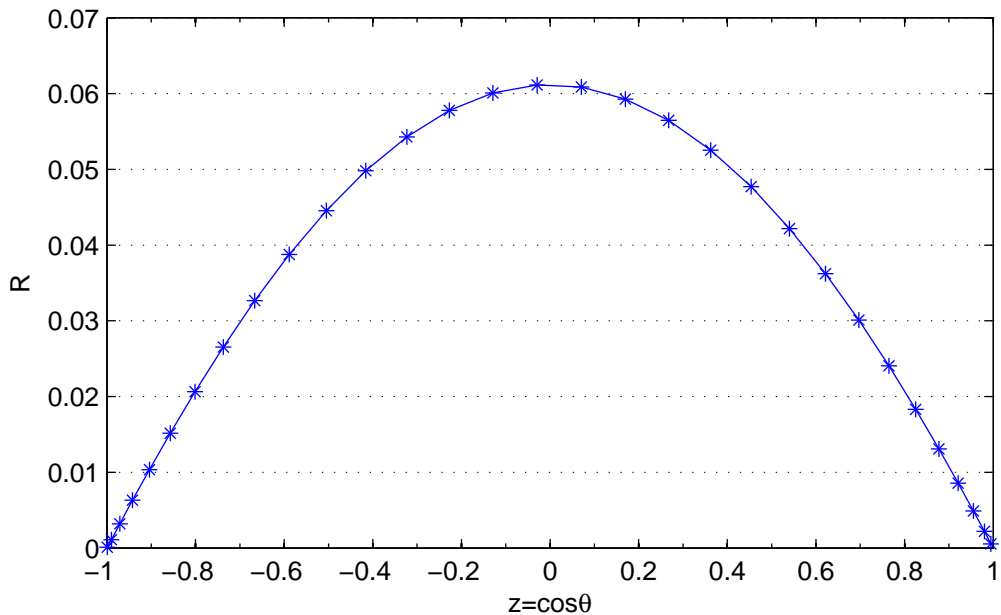}
\caption{The ratio $(\rm{d}\sigma-\rm{d}\sigma_{NC})/ \rm{d}\Omega$
at $\phi=\pi/2$(real line) and $\pi$($*$-dotted) for scattering
$\pi^+\pi^+ \rightarrow \pi^+\pi^+$. The reason of superposition for
the two different cases is that the noncommutative energy scale is
so large that the dependency of phase difference to $\phi$ become
unimportant.}\label{4}
\end{figure}

\section{$\pi^+\pi^-$ elastic scattering in NCSQED}
Now that we have completed our discussion of the process
$\pi^+\pi^+\rightarrow\pi^+\pi^+$, let us consider a different but
similar process $\pi^+\pi^-\rightarrow\pi^+\pi^-$. The lowest order
Feynman diagram is pictured in Fig.5.
\begin{figure}[h]
\centering
\includegraphics{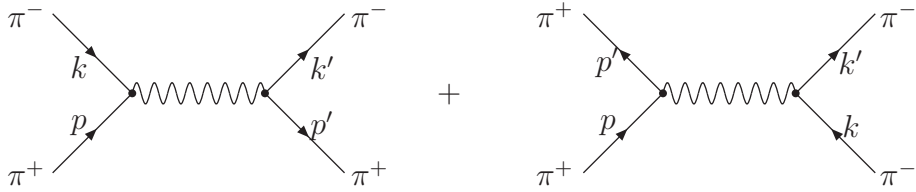}
\caption{The Feynman diagram of elastic scattering of
$\pi^+\pi^-\rightarrow\pi^+\pi^-.$} \label{5}
\end{figure}
Similar to the process of $\pi^+\pi^+\rightarrow\pi^+\pi^+$,
following the momentum labeling given in Fig.5, we can write down
the matrix element as
$$
    i{\mathcal {M}}_{+-}=-ie^2 \exp{\left[-{i\over 2}(p\wedge k+p'\wedge k')\right]}
    (p-k)^{\alpha}(k'-p')^{\beta}\frac{g_{\alpha\beta}}{q^2}
    ~~~~~~~
$$
\begin{equation}\label{22}
    +ie^2 \exp{\left[{i\over 2}(p\wedge p'+k\wedge k')\right]}
    (p+p')^{\alpha}(k+k')^{\beta}\frac{g_{\alpha\beta}}{q^2}.
\end{equation}
We can see that the $s$- and $t$-channal interference terms now pick
up a kinematic phase difference given by
\begin{equation}\label{23}
    \delta_{+-}={1\over 2}(p\wedge k+p'\wedge k')+{1\over 2}(p\wedge p'+k\wedge
    k').
\end{equation}
Neglecting the effect of noncommutativity of space-time, i.e.
$\theta_{0i}=0$, and using the formulas of (12-15,18-20), explicitly
we can obtain this phase difference
\begin{equation}\label{24}
    \delta_{+-}=\sqrt{ut}(\theta_{12}\cos{\phi}-\theta_{31}\sin{\phi})=-\delta_{++}.
\end{equation}
After a straightforward calculation we can obtain the squared
invariant matrix element
\begin{equation}\label{25}
    |{\mathcal {M}}_{+-}|^2=\bigg(\frac{4\pi\alpha}{s}\bigg)^2\bigg[(s+u)^2
    +(s+t)^2+2\cos{\delta_{+-}}(s+u)(s+t)\bigg].
\end{equation}
The differential cross section of this process having the same form
as of (21) can be expressed by
\begin{equation}\label{26}
    \bigg(\frac{\rm{d}\sigma_{+-}}{\rm{d}\Omega}\bigg)
    _{\rm{CM}}=\frac{|{\mathcal {M}}_{+-}|^2}{64\pi^2s^2}.
\end{equation}

\begin{figure}[h]
\centering
\includegraphics{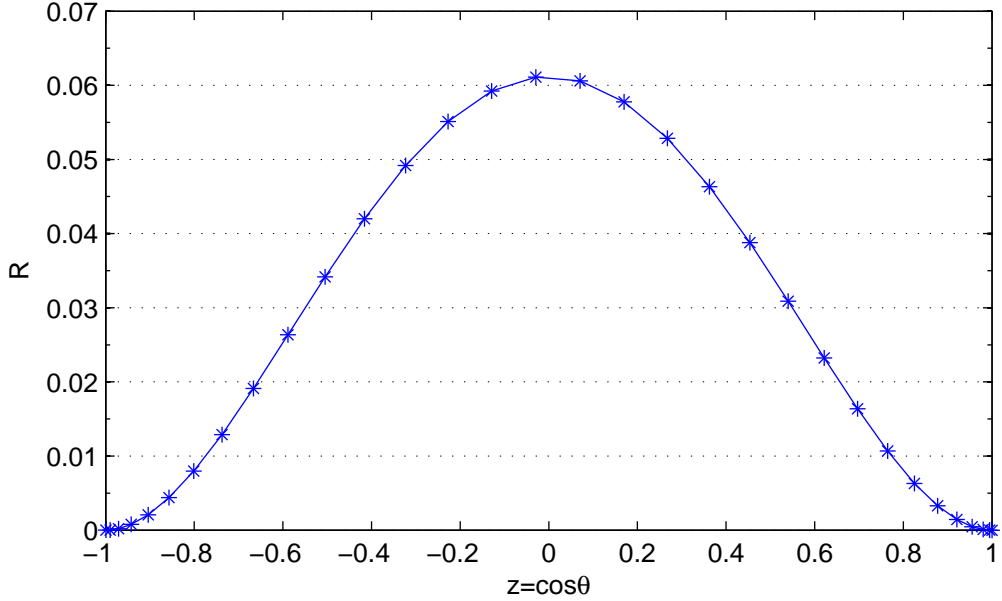}
\caption{The ratio $(\rm{d}\sigma-\rm{d}\sigma_{NC})/ \rm{d}\Omega$
at $\phi=\pi/2$(real line) and $\pi$($*$-dotted) for scattering
$\pi^+\pi^- \rightarrow \pi^+\pi^-$. As the same case and reason
there is a superposition for the two different cases. However, at
$\theta = 0,~\pi$ the change rates become slow compared to the
previous reaction.}\label{6}
\end{figure}

\section{Annihilation of $\pi^+\pi^-$ in NCSED}
In this section we will study the annihilation process of
$\pi^+\pi^-$ in NCSED. It is well known that some Feynman diagrams
have contributed a lot to this process. For examples, the 3-point
and 4-point photon functions contradict with 2-point photon
function. But it is a pity that all these diagrams contain the
photon-loop, we have to ignore those contributions made. The Feynman
diagrams of this annihilation up to tree level are draft out in
Fig.7.
\begin{figure}[h]
\centering
\includegraphics{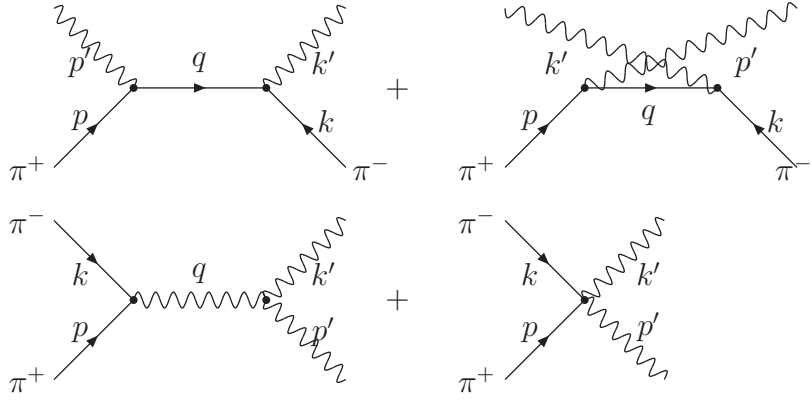}
\caption{The Feynman diagrams of annihilation of $\pi^+ \pi^-$}
\label{7}
\end{figure}
Due to the presence of the non-Ablelian-like coupling, we must be
careful in calculation of the cross section to ensure that the Ward
identities are satisfied so that the unphysical polarization states
are not produced. Meanwhile, we have to note that this process is
sensitive to space-time noncommutativity. So, the cross section will
not be invariant in all reference frames because of the violation of
Lorentz invariance which is a character of the noncommutative field
theory$^{[7]}$.

Using the Feynman rules given in section 1, we can now read the
invariant matrix element
$$
    i{\mathcal {M}^{\mu\nu}_{\rm{ap}}}= 2e^2\exp{\left[{i\over 2}p\wedge k\right]}
    (p+k)_{\rho}\frac{1}{q^2}
    \sin{(k'\wedge p')}[(k'-p')^{\rho}g^{\mu\nu}
    +(p'-q)^{\mu}g^{\nu\rho}+(q-k')^{\nu}g^{\rho\mu}]
$$
$$
    -ie^2 \exp{\left[-{i\over 2}(p\wedge p'+k\wedge k')\right]}
    \frac{(p+q)^{\mu}(k+q)^{\nu}}{q^2}
    ~~~~~~~~~~~~~~~~~~~~~~~~~~~~~~~~~
$$
$$
    -ie^2 \exp{\left[-{i\over 2}(p\wedge k'+k\wedge p')\right]}
    \frac{(p+q)^{\mu}(k+q)^{\nu}}{q^2}
    ~~~~~~~~~~~~~~~~~~~~~~~~~~~~~~~~~
$$
\begin{equation}\label{27}
    +2ie^2g_{\mu\nu}\cos{(p\wedge k)}\exp{[{i\over2}k'\wedge (p+
    k)]}.
    ~~~~~~~~~~~~~~~~~~~~~~~~~~~~~~~~~~~~~~~~~~~~
\end{equation}
If $\theta_{0i}=0$, i.e. only the effect of noncommutativity of
space-space is considered, the first line of (27) will vanish.
Accordingly, the phase difference of the $u$- and $t$-channel will
vanish. That is to say, there are no additional factors to the
interference terms between $u$- and $t$- channels, and the last term
will be a constant as the original case. Therefore, if there is some
influence arising from the modifications of noncommutative field
theory, the noncommutativity of space-time must be put into
consideration. As the Noncommutative Eletrodynamics, the
Noncommutative Scalar Eletrodynamics is also sensitive to the
noncommutativity of space-time.

As this is an unappealing theory in the case $\theta_{0i}\neq0$, we
don't intend to deduce the scattering cross section, although it is
possible to establish a consistent noncommutative theory in which
the scattering matrix is unitary via a new definition of the
Time-order in perturbation theory$^{[24]}$.

\section{Conclusion remarks }
In this paper we have analyzed various $2\rightarrow2$ high energy
scattering processes in the noncommutative scalar eletrodynamics. As
the modifications of eletrodynamics, both result in a
non-Abelian-like nature with 3- and 4-point photon self-couplings,
as well as momentum dependent phase factors appearing at each
possible vertex in noncommutative scalar QED. However, there are
some differences from the noncommutative QED due to the quadric
kinematic terms in the lagrangian, which can be realized explicitly
form the interaction lagrangian (9) or can be understood from the
Feynman rules in Fig.2. We have known that the interference of the
distinct channels plays a significant role. If there is only one
channel of reaction, the effect of noncommutativity will be hidden
for tree level diagram, although there are new Feynman rules of 3-
and 4-point photon self-couplings for the case $\theta_{0i}\neq 0$
as we discussed in the last section. In the numerical analysis, we
have parametrized the noncommutative relationship in terms of
NC-energy-scale $\Lambda_{\rm{NC}}$ and an antisymmetric unit matrix $c_{\mu\nu}$
,i.e. $\theta_{ij}=c_{ij}/\Lambda^2_{NC}$. We hope that these results can
be used, as a new approach, to test the non-commutativity of space.

\section{Acknowledgements}
This work was supported by the National Natural Science Foundation of China (10965006 and 11175053),
an open topic of the State Key Laboratory for Superlattices and Microstructures (CHJG200902), the
scientific research project in Shaanxi Province (2009K01-54) and  Natural Science Foundation of Zhejiang Province (Y6110470),

\end{document}